\documentclass[a4paper,11pt]{article}
\usepackage{graphicx}
\usepackage{xcolor}
\usepackage{amsmath,amsfonts,amssymb,amstext}
\usepackage[hidelinks]{hyperref}
\usepackage{hypcap}
\usepackage{multirow}
\usepackage{cancel}
\usepackage{empheq}
\usepackage{float}
\usepackage{cite}
\usepackage{subcaption}
\usepackage{authblk}
\usepackage{comment}
\usepackage{blindtext}
\usepackage{cases}
\usepackage{ulem}
\usepackage{epsfig}%
\usepackage[bottom]{footmisc}

\usepackage{epsfig}%
\usepackage[bottom]{footmisc}
\begin{document}
\date{}

\title{\centerline \bf Dynamical analysis of coupled  curvature-matter scenario in viable $f(R)$ dark energy models at de Sitter  phase}
\bigskip

\author[]{Anirban Chatterjee \thanks{iitkanirbanc@gmail.com \& anirban.chatterjee@gm.rkmvu.ac.in}}
\normalsize
\affil[]{Department of Physics\par
Ramakrishna Mission Vivekananda Educational and Research Institute\par Belur Math, Howrah 711202, West-Bengal, India}
\date{\today}
\maketitle

\begin{abstract}  
We explore the interaction between dark matter and curvature-driven dark energy within viable $f(R)$ gravity models, employing the phase-space analysis approach of linear stability theory. By incorporating an interacting term, denoted as $\mathcal{Q}=\alpha H \tilde{\rho}_{\rm m}\left(\frac{\kappa^2 }{3H^2}\rho_{\rm curv} + 1 \right)$, into the continuity equations of both sectors, we examine dynamics of two $f(R)$ gravity models that adhere to local gravity constraints and fulfill cosmological viability criteria. In de Sitter  phase, our investigation
reveals modifications of critical points compared to the conventional form, attributed to the introduced interaction. Through a comprehensive phase-space analysis, we illustrate the trajectories near critical points and outline the constraints on the phase space based on cosmological and cosmographic parameters.  Furthermore, we can also observe  stable behavior of this interacting system during the de Sitter phase of the universe.
\end{abstract}

\section{Introduction}

Observational evidence of redshift and luminosity distance for type Ia supernovae events  \cite{ref:Riess98, ref:Perlmutter} has been instrumental in validating the transition of the universe from a decelerated phase to its presently accelerated phase during late-time evolution. These evidences find support in the examination of temperature anisotropies within the cosmic microwave background data from the WMAP mission \cite{WMAP:2003elm, Hinshaw:2008kr} and the identification of baryon acoustic oscillations \cite{SDSS:2005xqv}.  In accordance with the general theory of relativity, the mechanism propelling accelerated expansion is attributed to a theoretical energy density component characterized by negative pressure, commonly known as dark energy. The inquiry into the origin and properties of dark energy, responsible for the ongoing cosmic acceleration, persists as a significant unsolved enigma in contemporary cosmology. Late-time acceleration of cosmos eludes explanation through the standard equation of state, denoted as $\omega = p/\rho$, where $p$ and $\rho$ signify the pressure and energy density of conventional universe constituents like radiation and matter. Evidently, there exists a necessity for an as-yet-unidentified component marked by negative pressure, yielding an equation of state $\omega < -1/3$, to elucidate the observed accelerated expansion. \\

Einstein initially introduced the $\Lambda g_{\mu\nu}$ cosmological constant term in his general relativity equations for a static universe. However, he later discarded it in favor of Hubble's evidence of an expanding universe. In the late 20th century, the term regained significance due to its potential to explain late-time cosmic acceleration, leading to the development of the $\Lambda$-CDM model, incorporating cold dark matter. Unfortunately, this model faces challenges such as the coincidence \cite{Zlatev:1998tr} and fine-tuning \cite{Martin:2012bt} problems, prompting the exploration of alternative dark energy models. One model category involves field-theoretic approaches to dark energy, modifying Einstein's field equations by introducing a scalar field as an additional universe component, distinct from matter and radiation. This category includes quintessence \cite{Peccei:1987mm,Ford:1987de,Peebles:2002gy,Nishioka:1992sg, Ferreira:1997au,Ferreira:1997hj,Caldwell:1997ii,Carroll:1998zi,Copeland:1997et} and $k-$essence \cite{Fang:2014qga, ArmendarizPicon:1999rj,ArmendarizPicon:2000ah,ArmendarizPicon:2000dh,ArmendarizPicon:2005nz,Chiba:1999ka,ArkaniHamed:2003uy,Caldwell:1999ew,Bandyopadhyay:2017igc,Bandyopadhyay:2018zlz,Bandyopadhyay:2019ukl,Bandyopadhyay:2019vdd,Chatterjee:2022uyw} models. Another model class focuses on altering the geometric aspects of Einstein's equations, particularly the Einstein-Hilbert action, to address late-time cosmic acceleration. These models, detailed in references \cite{fr1,fr2,fr3,fr4,fr5,fr6,fr7}, encompass $f(R)$ gravity, scalar-tensor theories, Gauss-Bonnet gravity, and braneworld models of dark energy. In many scenarios, the universe is assumed to be homogeneous and isotropic on large scales, as  described by the Friedmann-Lema\^itre-Robertson-Walker (FLRW) metric with a time-dependent scale factor, denoted as $a(t)$, and a curvature constant. Some approaches also consider an inhomogeneous universe, using a perturbed FLRW metric to represent the background spacetime. Dynamical systems techniques are powerful for studying cosmic evolution in both generic cosmological models and specific solutions. Recent applications include assessing the stability of scalar field dark energy models \cite{Chatterjee:2021ijw,Chatterjee:2021hhj,Hussain:2022osn,Bhattacharya:2022wzu,Hussain:2023kwk} and exploring $f(R)$ gravity \cite{fr8,fr9,fr10,fr11}. Comprehensive investigations into dynamical systems within $f(R)$ gravity theories are documented in various research endeavors, such as \cite{fr11,Samart:2021viu,Amendola:2006kh,Amendola:2006eh,Amendola:2007nt}. These studies aimed to categorize accurate cosmological models among different $f(R)$ gravity models. They contributed to developing `cosmologically viable' $f(R)$ gravity models that align with observed cosmic phenomena and meet local gravity constraints \cite{fr8,fr9,fr10,fr11}. These models offer cosmic acceleration without needing a cosmological constant or introducing dark energy as an additional field, relying on the curvature-driven dynamics described by the modified field equations with the function $f(R)$ of the Ricci scalar.  The main motivation for using $f(R)$ models is to explore the impact of dark energy during the late stages of the universe. This is achieved by modifying the Ricci scalar $R$ in the conventional Hilbert-Einstein Lagrangian density through a general function $f(R)$. Modified gravity, unlike other approaches, can emulate the characteristics of dark energy without introducing additional degrees of freedom. This eliminates the need for a cosmological constant ($\Lambda$-CDM), an extra scalar field (quintessence theory), an additional kinetic term with a scalar field ($k$-essence theory), or both a scalar field and Ricci scalar (scalar-tensor theory) in the Lagrangian. The reduced degrees of freedom in $f(R)$ gravity better align with observational datasets, and the dynamic system formulation becomes more compact, rendering the scenario autonomous. Notably, some $f(R)$ gravity models can explain late-time accelerated expansion while remaining consistent with local gravity constraints. By utilizing two viable $f(R)$ dark energy models, we have gained insights into the coupled curvature-matter system through a comprehensive dynamical analysis.\\

Here, we have investigated the impact of interacting curvature-driven dark energy and (dark) matter in viable $f(R)$ gravity models at de Sitter  phase.  Theoretical framework results from a modified geometric part and a matter component, ensuring minimal coupling with the geometry. The modified field equations, resembling standard field equations, involve the Einstein tensor and a total energy-momentum tensor. The latter comprises a curvature component determined by $f(R)$ and its derivatives, and a modified matter component derived by scaling the usual energy-momentum tensor from Einstein's equation. In FLRW spacetime background, conservation of total energy-momentum tensor leads to a continuity equation for the comprehensive fluid, including matter and curvature components. We introduce interactions between matter and curvature components with a source term $\mathcal{Q}$ in their continuity equations, representing the rate of energy exchange between them. The interaction term has been chosen for this work is, $\mathcal{Q}=\alpha H \tilde{\rho}_{\rm m}\left(\frac{\kappa^2 }{3H^2}\rho_{\rm curv} + 1 \right)$. In previous articles \cite{fr11, Samart:2021viu}, authors have opted for alternative interaction terms, such as those exclusively related to matter or a combination of matter and curvature. But here, the motivation behind choosing this particular form of interaction is to study the system's dynamics in the presence of both additive and multiplicative types of matter, matter-curvature interactions. Here, $\alpha$  indicates the coupling strength of the interaction term, and $\tilde{\rho}_{\rm m}$ and $\rho_{\rm curv}$ represent their respective energy densities.\\

Incorporating interactions between matter and curvature introduces the parameter $\alpha$ into this analytical framework, intertwined with parameters defining the $f(R)$ function. This coupling parameter significantly influences evolutionary dynamics in the matter-curvature coupled scenario. We use dynamical analysis to explore this interaction in viable $f(R)$ gravity models, setting up autonomous equations with dynamical variables. Identifying fixed points and assessing their stability through linear stability analysis is crucial for understanding aspects of cosmic evolution influenced by curvature-matter interactions. We focus on two $f(R)$ models, the generalized $\Lambda$-CDM model and the power-law model, both capable of driving cosmic acceleration while remaining cosmologically viable. To grasp the dynamics of the curvature-matter system in two $f(R)$ gravity models within the de Sitter  phase, we streamlined the approach from 3D to 2D variables and analyzed them using linear stability theory. From these phase-space plots, we can conclude the allowed region of each model at the de Sitter  phase in terms of stability,  cosmographic parameters, and the viability condition of critical energy density of matter.  Finally, we have drawn evolution plots depicting the behavior of 2D variables and observed their asymptotically stable behavior near stable critical points during the de Sitter phase of the universe.\\

The paper is organized as follows: In sec.\ [\ref{Sec:2}], we     have established a theoretical framework to address the interaction between the curvature and matter sectors in the background of a flat FLRW metric. In sec.\ [\ref{Sec:3}], we    have formulated the autonomous equations of the dynamical system, incorporating the interaction between matter and curvature-driven dark energy in the framework of $f(R)$ gravity. We have also discussed the two specific $f(R)$ models considered for this study.  In sec.\ [\ref{Sec:4}], we have examined this interacting scenario in the context of two different types of viable $f(R)$ models through the dynamical analysis approach and put some extra constraints from stability and other criteria.  Finally, We have summarized our conclusions in sec.\ [\ref{Sec:5}].\\

\section{Framework of curvature-matter interaction in $f(R)$ gravity}
\label{Sec:2}

Action of minimally coupled curvature-matter sector \cite{fr1,fr2,fr11}  in the context of $f(R)$ theory of gravity can be written as, 
\begin{eqnarray}
\mathcal{S} = \frac{1}{2\kappa^2}\int d^4x\sqrt{-g}f(R) + \int d^4x~\mathcal{L}_M(g_{\mu \nu},\phi _M)\,,
\label{eq:b1}
\end{eqnarray}
Here, $\kappa^2=8\pi G$ and $f(R)$ is any function of the Ricci scalar $R$, and ${\cal L}_{M}$ is the Lagrangian density for matter component of the universe, with $g$ as the determinant of the metric tensor $g{\mu\nu}$. In metric formalism, the modified field equation is obtained by varying this action with respect to the field $g_{\mu \nu}$ as,
\begin{eqnarray}
F(R) R_{\mu\nu}-\frac{1}{2}g_{\mu \nu}f(R) + g_{\mu \nu}\square F(R) - \nabla_{\mu}\nabla_{\nu} F(R)  = \kappa^2 \tilde{T}_{\mu\nu}^{(\rm M)}\,.
\label{eq:b2}
\end{eqnarray}
Here, $F(R)$ is the derivative of $f$ with respect to $R$, and $\tilde{T}_{\mu \nu}^{(\rm M)}$ is the stress-energy tensor for matter components given by
\begin{eqnarray}
\tilde{T}_{\mu \nu}^{(\rm M)} \equiv -{2 \over \sqrt{-g}}{\delta(\sqrt{-g}{\cal
L}_M)\over \delta(g^{\mu\nu})} \,.
\label{eq:b3}
\end{eqnarray}
Eq.\  (\ref{eq:b2}) may be reconstructed as \cite{fr1,fr11}
\begin{eqnarray}
 G_{\mu\nu} \equiv R_{\mu\nu}- \frac{1}{2}Rg_{\mu \nu} 
 = \kappa^2 \left(T_{\mu\nu}^{(\rm M)} + T_{\mu\nu}^{(\rm curv)}\right) 
 \equiv \kappa^2 T_{\mu\nu}^{(\rm tot)} \,,
\label{eq:b4}
\end{eqnarray}
 $G_{\mu\nu}$ depicts the Einstein tensor and
\begin{eqnarray}
\kappa^2 T_{\mu\nu}^{(\rm curv)} & \equiv & 
\frac{1}{F} 
\left[\frac{1}{2}(f-RF)g_{\mu \nu} +  \left(\nabla_{\mu}\nabla_{\nu}- g_{\mu \nu}\square \right)F 
\right] 
\label{eq:b5} \\
\mbox{and} \quad T_{\mu\nu}^{(\rm M)}
& \equiv &
\frac{1}{F} \tilde{T}_{\mu\nu}^{(\rm M)}\label{eq:b6} 
\end{eqnarray}
The reformulated field equation Eq. (\ref{eq:b4}) describes the  FLRW universe with a fluid represented by the comprehensive energy-momentum tensor $T_{\mu\nu}^{(\rm tot)} \equiv T_{\mu\nu}^{(\rm curv)} + T_{\mu\nu}^{(\rm M)}$. The influence of the function $f(R)$ is evident in the distinct components $T_{\mu\nu}^{(\rm curv)}$ and $T_{\mu\nu}^{(\rm M)}$, forming $T_{\mu\nu}^{(\rm tot)}$. The former, determined by $f(R)$ and its higher derivatives, vanishes for $f(R)=R$, acting as a fluid-equivalent portrayal of the curvature function. The latter modifies $\tilde{T}_{\mu\nu}^{(\rm M)}$ using a functional multiplier of $1/F$, reflecting gravity modifications from $f(R)$ in the revised stress-energy tensor for matter component. In a flat FLRW spacetime background, the unmodified stress-energy tensor $\tilde{T}_{\mu\nu}^{(\rm M)}$, representing matter as a perfect fluid, becomes diagonal with entries derived from combined energy densities $\tilde{\rho}_{\rm m}$. The matter component is considered non-relativistic. Consequently, the modified stress-energy tensor $T^{(M)}_{\mu\nu} = (1/F)\tilde{T}_{\mu\nu}^{(\rm M)}$ corresponds to a fluid with energy density $\rho_{\rm m} $ and zero pressure, where $ \rho_{\rm m} \equiv  \tilde{\rho}_{\rm m}/F$. Positivity of $\rho_m$ is maintained due to the inherent positivity of $\tilde{\rho_m}$ and $F$. The `00' and `$ii$' components of the modified field equation Eq. (\ref{eq:b4}) yield modified versions of the Friedmann equations under these considerations;
\begin{eqnarray}
 H^2 &=& \frac{\kappa^2 }{3}\Big{[}   \rho_{\rm m} +   \rho_{\rm curv}  \Big{]}\,,
\label{eq:b7}\\ 
 \dot{H} +  H^2 &=& - \frac{\kappa^2}{6}
\Big{[}     \rho_{\rm m}  + \rho_{\rm curv}  +   3  P_{\rm curv}\Big{]} 
\label{eq:b8}\,,
\end{eqnarray}
Here,  $H$ denotes the Hubble parameter, $ \rho_{\rm curv}$ and $P_{\rm curv}$ can be expressed as,
 \begin{eqnarray}
 \rho_{\rm curv} & \equiv &  
 \frac{1}{\kappa^2 F}\left(\frac{RF-f}{2}-3H\dot{R}F'\right)
\label{eq:b9}\\
 P_{\rm curv}& \equiv &\frac{1}{\kappa^2 F}\left(\dot{R}^2F'' + 2H\dot{R}F'+\ddot{R}F' +\frac{1}{2}(f-RF)\right)
\label{eq:b10}\,,
\end{eqnarray}
Where, $'$ signifies the derivative with respect to $R$. In a flat FLRW spacetime, the stress-energy tensor $T_{\mu\nu}^{\rm {curv}}$ for curvature-fluid is equivalent to that of an ideal fluid, with energy density and pressure represented by $\rho_{\rm curv}$ and $P_{\rm curv}$ as defined in Eqs.\ (\ref{eq:b9}) and (\ref{eq:b10}) from Eq.\ (\ref{eq:b5}).
We can also express Eq.\ (\ref{eq:b7}) as
\begin{eqnarray}
\Omega_{\rm m}  + \Omega_{\rm curv} &=& 1
\label{eq:b11}
\end{eqnarray}
Here $\Omega_{\rm m}$, and $\Omega_{\rm curv}$ are redefined density parameters by the following equations;
\begin{eqnarray}
\Omega_{\rm m} \equiv \frac{\kappa^2\rho_{\rm m}}{3H^2} = \frac{\kappa^2\tilde{\rho}_{\rm m}}{3FH^2}\,,\quad
 \Omega_{\rm curv} \equiv \frac{\kappa^2\rho_{\rm curv}}{3H^2}
\label{eq:b12}
\end{eqnarray}
%
%
Taking divergence in both sides of \eqref{eq:b4} and applying Bianchi's identity yields the conservation equation for the total stress-energy tensor as, 
\begin{eqnarray}
\nabla^{\mu}  T_{\mu\nu}^{\rm tot} &=& \nabla^{\mu}  \left(T_{\mu\nu}^{(\rm M)} + T_{\mu\nu}^{(\rm curv)}\right) =0 
\label{eq:b13}
\end{eqnarray}
In FLRW spacetime, \eqref{eq:b13} becomes a continuity equation for the inclusive fluid with energy density.
${\rho}_{\rm tot} \equiv    \rho_{\rm m} + \rho_{\rm curv} $
and pressure $P_{\rm tot} \equiv   P_{\rm curv}$
as 
\begin{eqnarray}
\dot{\rho}_{\rm tot}+3H{\rho}_{\rm tot}(1 + \omega_{\rm tot}) = 0\,,
\label{eq:b14}
\end{eqnarray}
Here, $\omega_{\rm tot}$ denotes the grand equation of state (EoS) parameter for the combined (matter and curvature) fluid. While the total stress-energy tensor $T^{(\rm tot)}{\mu\nu}$ remains conserved, individual components $T^{(\rm curv)}_{\mu\nu}$ and $T^{(M)}_{\mu\nu}$ may not be individually conserved. This allows for potential interactions between the matter and curvature sectors.\\

With these considerations, we express the conservation Eq. (\ref{eq:b13}) as $\nabla^{\mu} T_{\mu\nu}^{(\rm m)} = -\nabla^{\mu} T_{\mu\nu}^{(\rm curv)} \equiv - \mathcal{J}_{\nu} \neq 0$. In the context of FLRW space-time, non-conserving continuity equation of curvature-matter sectors can be written as,
\begin{eqnarray}
\dot{\rho}_{\rm curv}+3H\left(\rho_{\rm curv}+P_{\rm curv}\right)&=& \mathcal{Q} \label{eq:b15}\\
\dot\rho_{\rm m} +3H\rho_{\rm m} &=&-\mathcal{Q}
\label{eq:b16}
\end{eqnarray}
Where source term $\mathcal{Q}$ represents a time-varying function indicating the instantaneous rate of energy exchange between the curvature and matter sectors. Prior research has explored related avenues \cite{Samart:2021viu, fr11}, adopting interaction forms based solely on matter components or the multiplicative nature of both curvature and matter sectors. However, our study uniquely delves into both forms of interactions, $\mathcal{Q}=\alpha H \tilde{\rho}_{\rm m}\left(\frac{\kappa^2 }{3H^2}\rho_{\rm curv} + 1 \right)$, employing a comprehensive phase-space analysis method. We restrict our study only to the de Sitter  phase, with matter components minimally coupled to curvature sectors under a flat FLRW metric background. Here, $\alpha$ is the coupling strength of the interactions. This specific type of interaction incorporates degrees of freedom from both sectors and also ensures equality between matter and curvature in terms of critical density (where, $\frac{\Omega_{\rm m}}{\Omega_{\rm curv}}=r_{\rm mc} \sim 1$), depending on the strength of the coupling parameter ($\alpha$).\\

Imposing $\rho_{\rm curv} > 0$ assigns the concept of `energy density' to it, placing constraints on $f(R)$ models to satisfy $6H\dot{R}F' < RF - f $ as seen in Eq.\ (\ref{eq:b9}). This constraint also impacts the coupling parameter $\alpha$ through its connection with source term $Q$ in Eq.\ (\ref{eq:b15}). Under this condition, each modified density parameter $\Omega_{\rm m}$, and $\Omega_{\rm curv}$ stays positive and collectively follows the constraint in Eq.\ (\ref{eq:b11}). This allows us to explore the parameter space for curvature-matter interaction by varying $\Omega_{\rm m}$ within the range $0 \leqslant \Omega_{\rm m} \leqslant 1$, leading to constraints on the model parameters. Additional constraints arise from the non-phantom and accelerating nature of the grand equation of state (EoS) parameter, along with cosmographic parameters (deceleration and jerk) and the critical energy density ratio of matter-to-curvature components. Further details on these constraints will be provided in the following section. \\

\section{Dynamical analysis of interacting matter-curvature  scenario in two types of $f(R)$ models}
\label{Sec:3}
Dynamical analysis is employed to investigate curvature-matter interaction in cosmologically viable $f(R)$ gravity models. We can express autonomous equations in terms of dynamically relevant variables, laying the groundwork for applying dynamical analysis techniques in our context. Fundamental dynamical variables are,
\begin{eqnarray}
X_{1}= - \frac{\dot F}{HF}, \quad  X_{2}= - \frac{f}{6FH^2}, \quad X_{3}=\frac{R}{6H^2}
\label{eq:d0}
\end{eqnarray}
Introducing the dimensionless parameter $N = \ln a$ and using the defined dynamical variables ($X_i$'s), Eqs.\ (\ref{eq:b9}), (\ref{eq:b15}), and (\ref{eq:b16}) become a set of four autonomous equations, forming a 3-D dynamical system.
\begin{eqnarray}
\frac{dX_{1}}{dN}&=& -1 - X_{1}X_{3}- X_{3} - 3X_{2}  + X_{1}^2 \nonumber\\
&&\qquad +\ \alpha \left(1- X_{1} - X_{2} - X_{3} \right)
 \left(1+ X_{1} + X_{2} + X_{3}\right)  \label{eq:d1} \\
\frac{dX_{2}}{dN} &=& \frac{X_{1}X_{3}}{m}-X_{2}\left(2X_{3}-4-X_{1}\right)  \label{eq:d2}\\
\frac{dX_{3}}{dN} &=& -\frac{X_{1}X_{3}}{m} - 2X_{3}\left( X_{3}-2\right) \label{eq:d3} \,,
\end{eqnarray}

Here, $m$ and $r$ parameters can be defined as, 
\begin{eqnarray}
m &\equiv & \frac{d \ln F}{d \ln R}= \frac{R F' }{F}\nonumber\\
r & \equiv & -\frac{d \ln f}{d \ln R}= -\frac{R F}{f} = \frac{X_3}{X_2}\,.
\label{eq:d5}
\end{eqnarray}
The dimensionless parameters $m$ and $r$ depend on $R$ through $f(R)$, enabling us to express $m$ as a function of $r$, denoted as $m = m(r)$. Each specific functional relation $m = m(r)$ corresponds to a unique class of $f(R)$ models. With $\alpha$ sets to zero, the equations reduce to autonomous equations for a scenario without interactions between curvature and matter sectors. \\

Eqs.\ (\ref{eq:b11}) and (\ref{eq:b12}) impose constraints on the dynamical variables in Eq.\ (\ref{eq:d0}) as,
\begin{eqnarray}
\Omega_{\rm m}   = 1 - X_{1} - X_{2} - X_{3} 
\label{eq:d6}
\end{eqnarray}
We can further express  $\Omega_{\rm curv}$ as, 
\begin{eqnarray}
\Omega_{\rm curv}  =  1 - \Omega_{\rm m}  =  X_{1} + X_{2} + X_{3}\,.
\label{eq:d7}
\end{eqnarray}
Furthermore, the grand EoS parameter ($\omega_{\rm tot}$) can be written as, 
\begin{eqnarray}
\omega_{\rm tot} &=& -1 - \frac{2\dot{H}}{3H^2} = -\frac{1}{3}(2X_{3}-1)
\label{eq:d8}
\end{eqnarray}
The ratio of critical energy density of matter to curvature, denoted as $r_{mc} \equiv \Omega_{\rm m}/\Omega_{\rm curv}$, along with two cosmographic parameters - deceleration $(q \equiv - a\ddot{a}/\dot{a}^2)$ and jerk $(j \equiv \dddot{a}/aH^3)$ - serve as valuable indicators for assessing various aspects of cosmic dynamics, mainly in late-time scenario. In this matter-curvature interaction scenario, the expressions for the three parameters (\(r_{mc}, q, j\))  can be expressed in terms of the dynamical variables ($X_i$'s) as,  
\begin{eqnarray}
 r_{mc} &\equiv & \frac{\Omega_{\rm m}}{\Omega_{\rm curv}} 
 = \frac{1 - X_1 - X_2 - X_3}{X_1 + X_2 + X_3} \label{eq:d9}\\
q &\equiv & - \frac{a\ddot{a}}{\dot{a}^2} = -1 - \frac{\dot{H}}{H^2} = 1 - X_3  \label{eq:d10} \\
j &\equiv & \frac{\dddot{a}}{aH^3} = -\frac{X_1 X_3}{m}+2 (1 - X_3)^2+(1 - X_3)-2 X_3 (X_3-2)  \label{eq:d11}
\end{eqnarray}

The fixed points in the 3-D dynamical system, where $dX_i/dN = 0$ ($i=1,2,3$), are crucial for understanding cosmic evolution driven by curvature-matter interactions. Linear stability analysis, involving a first-order Taylor expansion, is used to assess stability around these points. The Jacobian matrix, $\rm J = \vert\vert {\partial f_i}/{\partial x_j}\vert\vert$ aids in determining stability, with asymptotic stability (unstable) indicated when all eigenvalues have negative (positive) real parts. A saddle point is identified if any pair of eigenvalues has a relative opposite sign in their real parts. If any eigenvalue approaches zero, center manifold theory is required for a more insightful exploration of these critical points characteristics.\\


This entire framework of dynamical analysis is based on $f(R)$-driven dark energy models, considering interactions between curvature and matter sectors. We selected specific models capable of inducing cosmic acceleration while remaining cosmologically viable. In the metric formalism, any viable $f(R)$ function must meet stringent conditions discussed in \cite{Faraoni:2008mf, Tsujikawa:2010sc}. Below, we list and provide the rationale for these conditions. We examine the autonomous system, focusing on two viable scenarios: The first involves a $f(R)$ yielding a constant $m$, while the second pertains to a $f(R)$ where $m$ is a function of $r$, denoted as $m(r)$. These scenarios are labeled as (I) and (II), with their specifications discussed below. 

\begin{itemize}
\item[(I)]  

In the modified  gravity model, $f(R) = (R^b - \Lambda)^c$ with $(c \geqslant 1, bc \approx 1)$, a constant $m$ scenario arises. Explored without considering matter-curvature interactions in \cite{fr10, Amendola:2007nt}, this model extends the $\Lambda$-CDM model, termed `generalized $\Lambda$-CDM model'. Constrained to $c \geqslant 1$ and $bc \approx 1$, it converges to the $\Lambda$-CDM model with $m=0$, ensuring viable cosmological evolution. The parameters $m$ and $r$ are given by $m= \frac{(bc - 1)R^b - b\Lambda + \Lambda}{R^b - \Lambda}$, $r=-\frac{bcR^b}{R^b-\Lambda}$, and $m$ can be expressed as $m(r)=\left(\frac{1-c}{c}\right)r+b-1$. A dynamical analysis (see \cite{fr10, Amendola:2007nt} for details) indicates stable points where $r = -1 - m$, resulting in $m=-1+bc$, a constant value.

\item[(II)] 
Power law model considers $m$ as a specific function of $r$ given by $m(r) = \frac{n(1+r)}{r}$, attainable with the power law form $f(R) = R - \gamma R^n$ ($\gamma>0$, $0<n<1$). This scenario accurately describes cosmic evolution within the framework of non-interacting curvature-matter scenarios, meeting the conditions for cosmological viability within the specified range of $\gamma$ and $n$, as extensively discussed in \cite{fr10,Li:2007xn}. The corresponding $f(R)$ form yields $m=\frac{\gamma(n-1)nR^n}{R - \gamma n R^n}$ and $r=\frac{R - \gamma n R^n}{R-\gamma R^n}$, with the elimination of $\gamma$ leading to the form $m(r) = \frac{n(1+r)}{r}$.

\end{itemize}  

In our study, the selection of the two cases: (I) $m=-1+bc=$ constant and (II) $m = m(r) = \frac{n(1+r)}{r}$ aims to investigate implications of the mentioned $f(R)$-models in the presence of matter-curvature interactions at the de Sitter  phase.\\

From the formulation of the dynamical variables in Eq. (\ref{eq:d0}), it is evident that both $X_1$ and $X_2$ are contingent on the chosen $f(R)$ models. In contrast, $X_3$ remains independent of the $f(R)$ selection and assumes a pivotal role in defining the grand equation of state parameter. Given our primary objective of investigating the dynamics of curvature-matter interaction in viable $f(R)$ dark energy models specifically during the de Sitter  phase, we have kept the variable $X_3$ fixed. This action reduces the phase-space dimension from 3-D to 2-D, focusing exclusively on a constant $X_3$ plane. The grand Equation of State (EoS) parameter at the de Sitter  phase, as indicated in Eq. \eqref{eq:d8}, yields a value of $-1$, which consequently fixes the $X_3$ variables at 2. Simultaneously, the deceleration parameter ($q$) in Eq. \eqref{eq:d10} is also negative unity at this point. Subsequently, we have established a 2-D phase space by varying $X_1$ and $X_2$ for specific benchmark values of $X_3$ (set to 2), where the system behaves akin to a de Sitter  solution.

\section{Phase-space analysis of matter-curvature coupling in modified $f(R)$ gravity models}
\label{Sec:4}
In both models, we identify fixed points of the autonomous system (see in sec.\ [\ref{Sec:3}]) governing curvature-matter interactions in $f(R)$ gravity at de Sitter  phase. Obtained by solving $dX_k/dN = 0$ ($k=1,2$), using $m = -1+bc$ for model I and $m = \frac{n(1+r)}{r} = \frac{n(X_2+2)}{2}$ for model II (where, $X_3=2$ at de Sitter  phase). Coupling parameter $\alpha$ with $(b,c)$ for model I and $(n)$ for model B are denoted as the free parameters of this interacting picture. Critical density parameters ($\Omega_{\rm m}$, $\Omega_{\rm curv}$) at fixed points $(X_i)$ are evaluated using Eqs.\ (\ref{eq:d6}), (\ref{eq:d7}). Positive modified density parameters, subject to the constraint in Eq.\ (\ref{eq:b11}), determine cosmological significance. Describing fixed points and their stability references relevant constraints within the model parameter space. Value of density parameters at each fixed point provides insights into characterizing the associated de Sitter  phase. In the following subsections, we have thoroughly examined interacting curvature-matter framework in the context of two types of $f(R)$ models at de Sitter  phase through linear stability analysis theory. By fixing $X_3 = 2$, we've analyzed the fixed points behavior with variables $X_1$ and $X_2$ and illustrated their dynamics in phase space.

\subsection{Analysis of generalized $\Lambda$-CDM model:}
We have initiate our investigation by employing the generalized $\Lambda$-CDM model to scrutinize the dynamic behavior of the coupled system. By exploiting the set of autonomous eqautions (\ref{eq:d1}), (\ref{eq:d2}) and also setting $X_3=2$, a total of four critical points have been found.  We have tabulated the critical points in tab.[\ref{Tab:T1}].\\
\begin{table}[h]
\centering
 \begin{tabular}{|c|c|c|}
\hline
\multicolumn{3}{|c|}{Critical points  (at $X_3 = 2$)} \\ 
\hline 
Critical Points & $X_1$ & $X_2$ \\
\hline
\hline
$ P_1 $ & $ 0 $ & $ -1 $   \\
\hline 
$ P_2 $ & $ 0 $ & $ -\frac{3 (\alpha +1)}{\alpha } $   \\
\hline
$ P_{3\mp} $ & 
$ \frac{-8 \alpha -4 \alpha  \text{bc}^2-2 \text{bc}^2+12 \alpha  \text{bc}+4 \text{bc}-2}{(2 (-1 + 2 bc - bc^2 + \alpha - 2 bc \alpha + bc^2 \alpha))} $ 
& $ -\frac{2}{\text{bc}-1} $   \\
& $ \mp \frac{1}{2 \left(\alpha +\alpha  \text{bc}^2-\text{bc}^2-2 \alpha  \text{bc}+2 \text{bc}-1\right)}\Big{[}-4 \left(\alpha +\alpha  \text{bc}^2-\text{bc}^2-2 \alpha  \text{bc}+2 \text{bc}-1\right)$& \\
& $\left(15 \alpha +3 \alpha  \text{bc}^2+3 \text{bc}^2-14 \alpha  \text{bc}-12 \text{bc}+9\right)$&  \\
&  $+\left(8 \alpha +4 \alpha  \text{bc}^2+2 \text{bc}^2-12 \alpha  \text{bc}-4 \text{bc}+2\right)^2\Big{]}^{1/2}$ &\\

\hline
\end{tabular}
\caption{Critical points of generalized $\Lambda$-CDM model at de Sitter  phase}
\label{Tab:T1}
\end{table}

\begin{figure}[H]
\centering
\includegraphics[width=.495\textwidth,height=5.5cm]{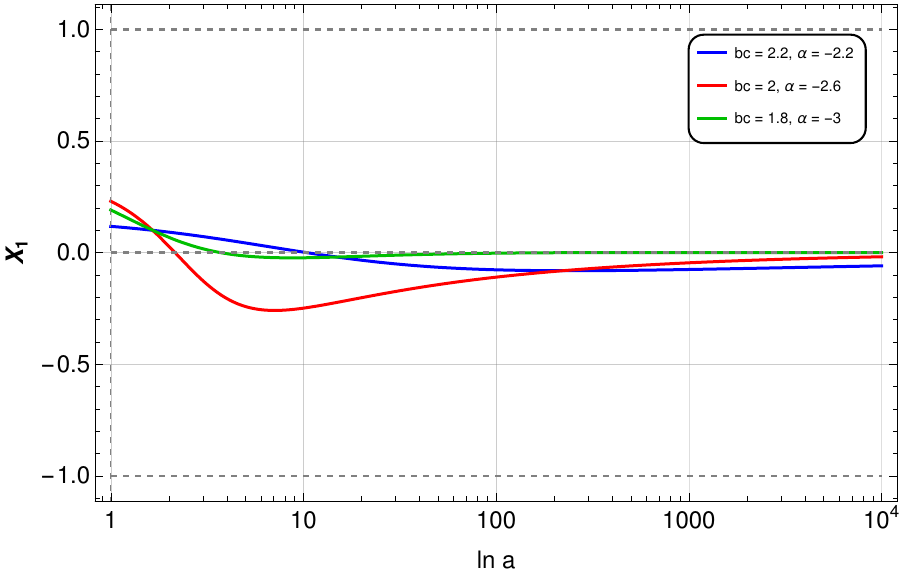}\hfill%
\includegraphics[width=.495\textwidth,height=5.5cm]{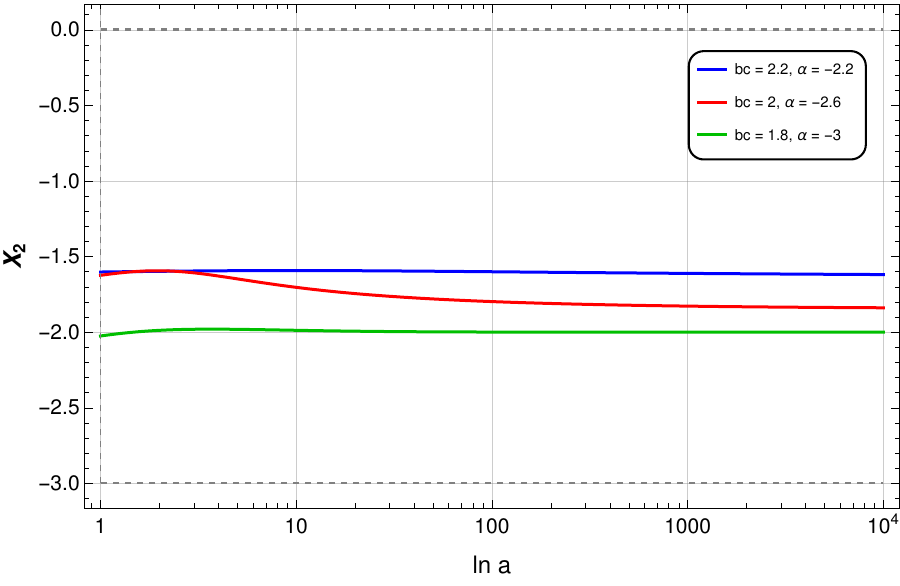}\hfill
\caption{ Evolution of $X_1$ and $X_2$ in generalized $\Lambda$-CDM model at three different benchmark values during the de Sitter phase.}
\label{fig:A3}
\end{figure}

Further details about the attributes of these critical points are discussed below,

\begin{itemize}
\item $P_1$ Point: This critical point remains model and coupling-parameter independent. The critical matter density ($\Omega_m$) at this point is consistently zero, making it a curvature-dominant critical point. Additionally, the ratio of critical matter to curvature density at this specific point is always zero. Although it exhibits a positive unit jerk in late times, the positive eigenvalues render this point consistently unstable.

\item $P_2$ Point: The characteristics of this critical point solely depend on the coupling parameter. The critical matter density ($\Omega_m$) and the ratio of critical matter to curvature density ($r_{\rm mc}$) at this point can be expressed in terms of the coupling parameter $\alpha$ as $\frac{3 (\alpha +1)}{\alpha }-1$ and $\frac{\frac{3 (\alpha +1)}{\alpha }-1}{2-\frac{3 (\alpha +1)}{\alpha }}$, respectively. Similar to the previous critical point, this point consistently exhibits a positive jerk of unity. The conditions for stability, critical matter density, and a positive ratio of critical matter to curvature density impose a common constraint in the parameter space spanned by $bc$ and $\alpha$, as illustrated in fig. [\ref{fig:A1}].\\
\vspace{-0.5cm}
\begin{figure}[h]
\centering
\includegraphics[width=.62\textwidth,height=8.1cm]{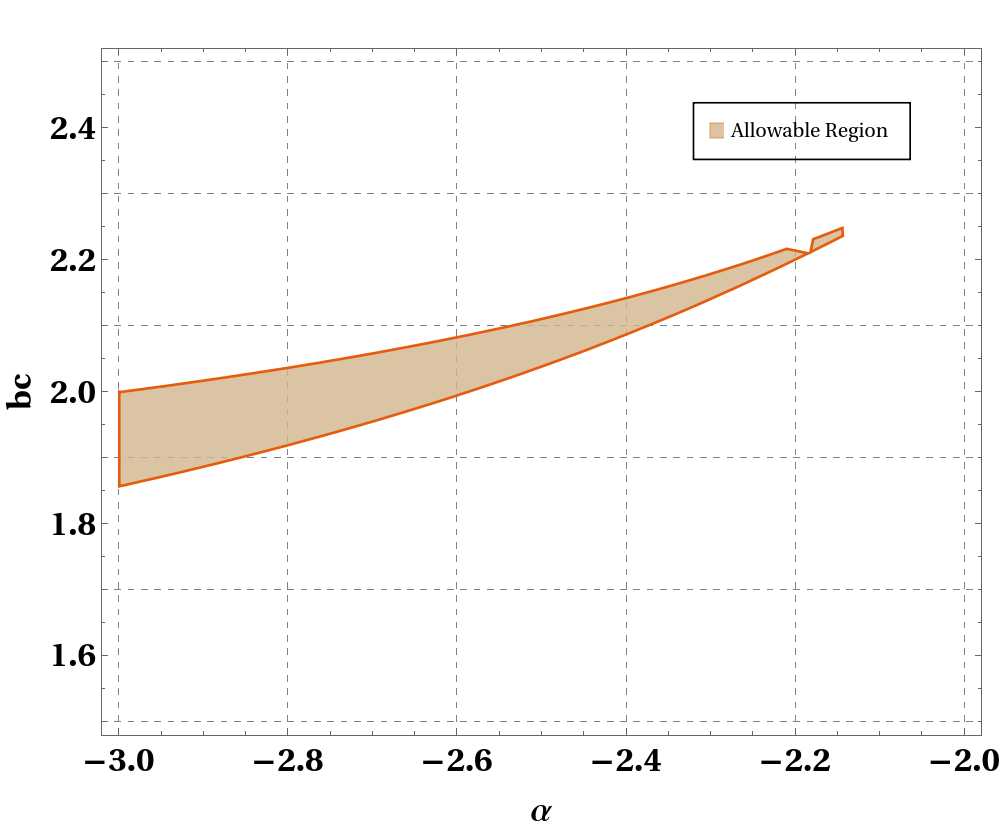}
\caption{Parameter space of generalized $\Lambda$-CDM model at $P_2$ in de Sitter  phase}
\label{fig:A1}
\end{figure}

From the shared permissible region defined by stability conditions ($r_{\rm mc}\geq 0$, $0 \leq \Omega_m \leq 1$), we selected three specific values for the parameters $bc$ and $\alpha$ to generate a phase-space plot involving variables $X_1$ and $X_2$. As shown in fig.[\ref{fig:A2}], we illustrate the behavior of both critical points, $P_1$ (independent of model parameters) and $P_2$ (dependent on model parameters). Across all three plots, the phase-space behavior of $P_1$ exhibits a repeller-type nature, with trajectories being repelled from $P_1$ and attracted towards $P_2$, establishing $P_2$ as a global attractor point. The emergence of $P_2$ is attributed solely to the presence of interaction. In the phase plot, the $X_2$ coordinate of $P_2$ is observed to be dependent on the coupling parameter ($\alpha$), shifting towards the negative direction for large negative values of the coupling parameter. The yellow band in the phase space indicates critical matter density between 0 and 1, with a positive jerk depicted in the magenta-marked region. The central green line within the yellow band signifies the epoch when critical matter and curvature density are the same. At this point, the $r_{\rm mc}$ parameter attains a value of one. In the context of a significantly negative coupling parameter ($\alpha \sim -3$), the attractor point $P_2$ is situated at the border of the yellow band. The analysis of the phase space at the de Sitter  point leads to the conclusion that curvature-matter interaction results in a critical point that satisfies all conditions and yields a stable attractor critical point for the system.
 
\item $P_{3\mp} $ Points: Only these points can be expressed regarding both model and coupling parameters. However, they fail to meet the specified physicality conditions and yield unstable critical points through linear stability analysis. The energy density and cosmographic parameters result in non-physical values at these points within the accessible region of model parameters identified from the $P_2$ point. Due to this restriction, these points cannot be included in the phase space diagram in fig.[\ref{fig:A2}].

\begin{figure}[H]
\centering
\includegraphics[width=.45\textwidth,height=7.9cm]{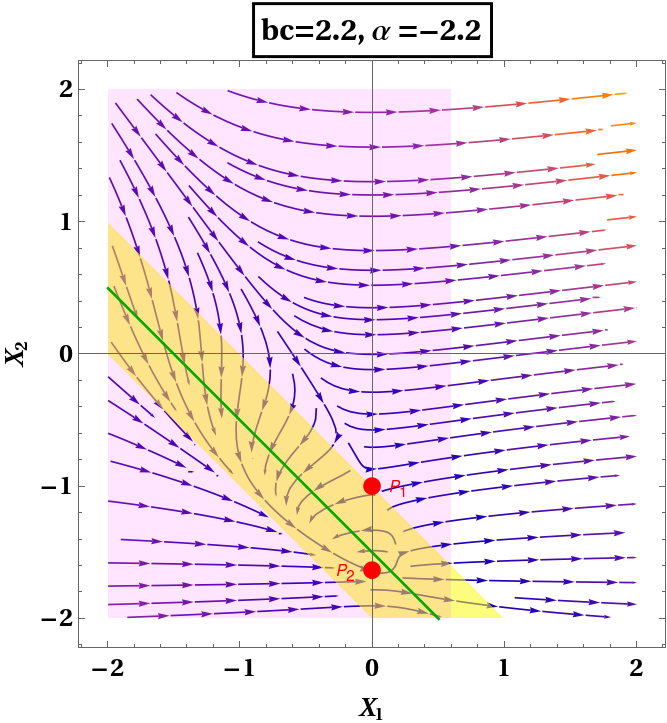}\hfill%
\includegraphics[width=.45\textwidth,height=7.9cm]{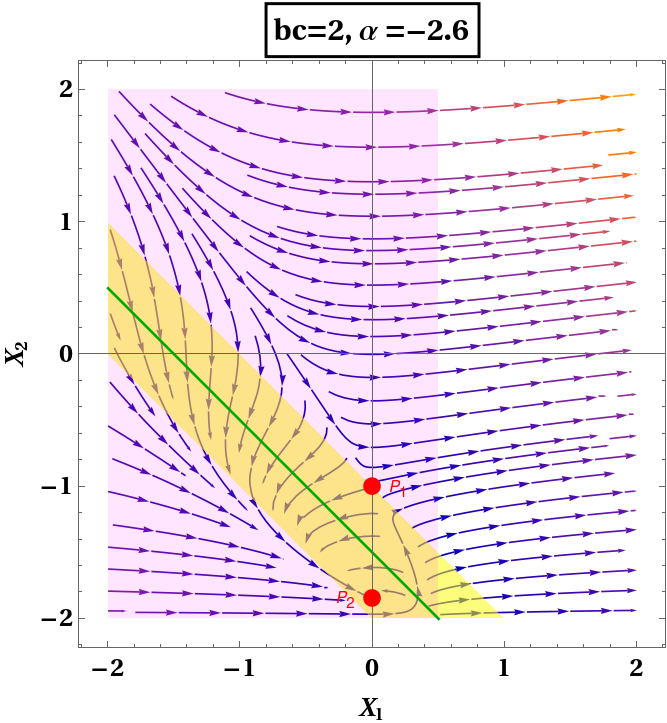}\hfill
\includegraphics[width=.45\textwidth,height=7.9cm]{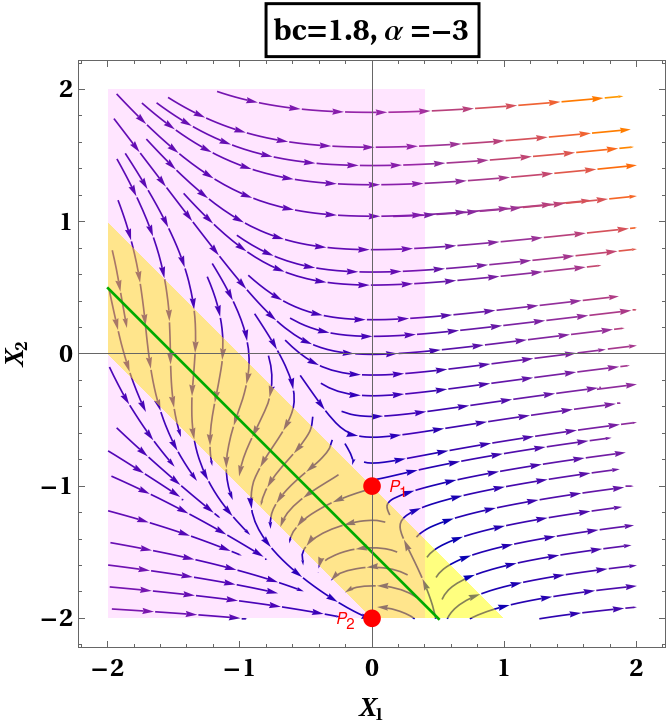}
\vspace{-3mm}
\caption{Phase space trajectories of generalized $\Lambda$-CDM model at different benchmark points. Critical points $P_1$ and $P_2$ are identified as repeller and attractor, respectively. The magenta region signifies positive jerk, and the yellow band represents $0 \leq \Omega_m \leq 1$ with $r_{\rm mc} \geq 0$. The central green line within the yellow band indicates the point where $r_{\rm mc}=1$.}
\label{fig:A2}
\end{figure}

In fig. [\ref{fig:A3}], we have illustrated the evolutionary dynamics of the dynamical variables $X_1$ and $X_2$ with the newly defined time parameter ($\ln a$), while holding the value of $X_3$ constant at 2, capturing the behavior of these variables during the de Sitter phase of the universe. The chosen values for the  coupling and model parameters lie within the permissible range of the parameter space defined by $\alpha$ and $bc$ for the generalized $\Lambda$-CDM case, and these values serve as benchmark points illustrated in fig. [\ref{fig:A1}]. As our main emphasis is on studying the dynamic characteristics during the de Sitter phase, highlighting the evolution from late-time to the distant future. Therefore, we are not concerned with demonstrating the highly unstable behavior of the dynamical variables during the early epochs. In the left panel, it is apparent that the $X_1$ variable converges toward a value near zero, while in the right panel, the $X_2$ variable saturates at three distinct values contingent on the coordinates of the critical points, as depicted in the three subfigures of fig.[\ref{fig:A2}]. Importantly, both variables exhibit asymptotically stable behavior, indicating a stable acceleration in the late-time evolution with a specific choice of coupling and model parameters, all of which adhere to physical viability conditions.

\end{itemize}

\subsection{Analysis of power-law model:}
We commence our exploration by utilizing the power-law model to examine the dynamic characteristics of the autonomous  system at de Sitter  phase. By analyzing the set of autonomous equations (\ref{eq:d1}) and (\ref{eq:d3}) while fixing $X_3$ to 2, a total of four fixed points have been identified, which are given in tab.[\ref{Tab:T2}].
\begin{table}[H]
\centering
 \begin{tabular}{|c|c|c|}
\hline
\multicolumn{3}{|c|}{Critical points  (at $X_3 = 2$)} \\ 
\hline 
Critical Points & $X_1$ & $X_2$ \\
\hline
\hline
$ P_1 $ & $ 0 $ & $ -1 $   \\
\hline 
$ P_2 $ & $ 0 $ & $ -\frac{3 (\alpha +1)}{\alpha } $   \\
\hline
$ P_{3\mp} $ & \scriptsize $2 (\alpha  n-n)\Big{[}-2 \alpha  \left(-\sqrt{n^2-4 n}-n\right)-4 \alpha  n-2 n$
& $ \frac{-\sqrt{n^2-4 n}-n}{n} $   \\
& & \\
&\scriptsize $\mp 2 \sqrt{n^2-3 n \sqrt{n^2-4 n}-4 \alpha  n+3 \alpha  n^2 + \alpha ^2 \left(n^2-4 n\right)-\alpha  n \sqrt{n^2-4 n}+4 \alpha ^2 n} \Big{]}$ & \\
& & \\
\hline
$ P_{4\mp} $ & \scriptsize $2 (\alpha  n-n)\Big{[}-2 \alpha  \left(-\sqrt{n^2-4 n}-n\right)-4 \alpha  n-2 n$
& $ \frac{-\sqrt{n^2-4 n}-n}{n} $   \\
& & \\
&\scriptsize $\mp 2 \sqrt{n^2-3 n \sqrt{n^2-4 n}-4 \alpha  n+3 \alpha  n^2 + \alpha ^2 \left(n^2-4 n\right)+\alpha  n \sqrt{n^2-4 n}+4 \alpha ^2 n} \Big{]}$ & \\
& & \\
\hline
\end{tabular}
\caption{Critical Points of power-law model at de Sitter  phase}
\label{Tab:T2}
\end{table}

Like the previous model, here we also employ linear stability theory to analyze the dynamical behavior of power-law model. This involves determining the Jacobian matrix and subsequently evaluating the stability of critical points. Further details regarding the properties of these critical points are discussed below,
\begin{itemize}
\item $P_1$ Point: This specific critical point remains independent of model parameters. The critical matter density is consistently zero at this juncture, indicating the dominance of curvature-driven dark energy. The stability condition of eigenvalues, $r_{\rm mc}\geq 0$ and $0 \leq \Omega_m \leq 1$ establish an allowed region, which is expressed in terms of the coupling parameter $\alpha$ and model parameter $n$. In fig.  [\ref{fig:B1}], we illustrate the permissible region of this power-law model during the de Sitter  phase corresponding to the $P_1$ point. The parameter space nature is divided into two regions based on the model parameter $n$, namely (i) $0<n<1$ and (ii) $n>1$. In the first region, the accessible range of coupling strength ($\alpha$) initiates from a numerical value of approximately $5.5$, while in the second case, the allowed range of coupling parameter ($\alpha$) starts from a small negative value ($\sim -0.5$). The coupling parameter $\alpha$ plays a crucial role in constraining $n$ in terms of stability and other physical conditions, as mentioned earlier.\\

\begin{figure}[H]
\centering
\includegraphics[width=.7\textwidth,height=9.5cm]{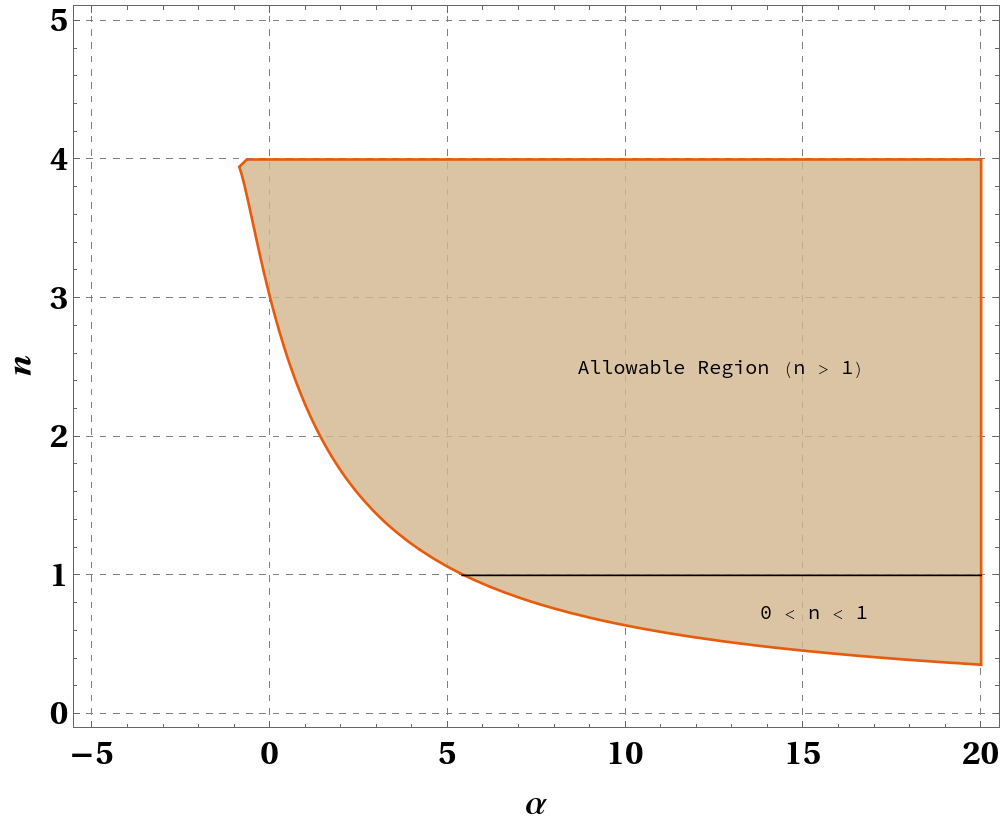}
\caption{Parameter space of $n$ vs. $\alpha$ of power-law model at $P_1$ in de Sitter  phase}
\label{fig:B1}
\end{figure}

\item $P_2$ Point: While this critical point is expressed as a function of the coupling strength $\alpha$, we are unable to identify any stable critical point. Due to its tendency to exhibit unstable behavior, we refrain from presenting detailed information on energy density and other parameters for this specific point.\\

We have selected two benchmark values for model parameters ($n, \alpha$) from each region in the parameter space shown in fig. [\ref{fig:B1}] and examined the 2-D phase-space behavior, as depicted in fig. [\ref{fig:B2}]. In the first region, we have chosen lower values for both coupling strength ($\alpha$) and model parameter ($n$) to observe the phase-space trajectories in the $X_1$-$X_2$ plane. In the power-law model, we have found that the value of $m = n (X_2 + 2)/2$, and when $X_2 = -X_3= -2$, $m$ becomes zero, rendering dynamical Eq. (\ref{eq:d2}) undefined at the de Sitter  solution. For this issue, we have identified a disjoint phase-space behavior marked by a black dotted line, where the dynamical variable $X_2 = -X_3= -2$. We have chosen benchmark values from the accessible range of $n, \alpha$ for the model parameter-independent critical points $P_1$, which also produced values for point $P_2$, plotted in the same phase space. All trajectories are repelled from the point $P_2$ and attracted towards $P_1$ making it into the attractor. Higher positive values of both parameters ($n, \alpha$) influence the movement of the critical point $P_2$ along the negative $X_2$-axis. Similar to the previous model, we marked a magenta region for a positive jerk parameter and a yellow band for the region where critical matter density is between 0 and 1, as well as the ratio of critical matter to curvature density being greater than zero. The middle green line in the yellow band indicates the time when critical matter and curvature density are equal. At this juncture, the $r_{\rm mc}$ parameter reaches a value of one. In all plots, the attractor point $P_1$ satisfies all conditions and lies inside the common intersecting region of the magenta and yellow bands. Similar behavior has been observed in the other two cases where $n>1$.
\begin{figure}[H]
\centering
\includegraphics[width=.45\textwidth,height=8cm]{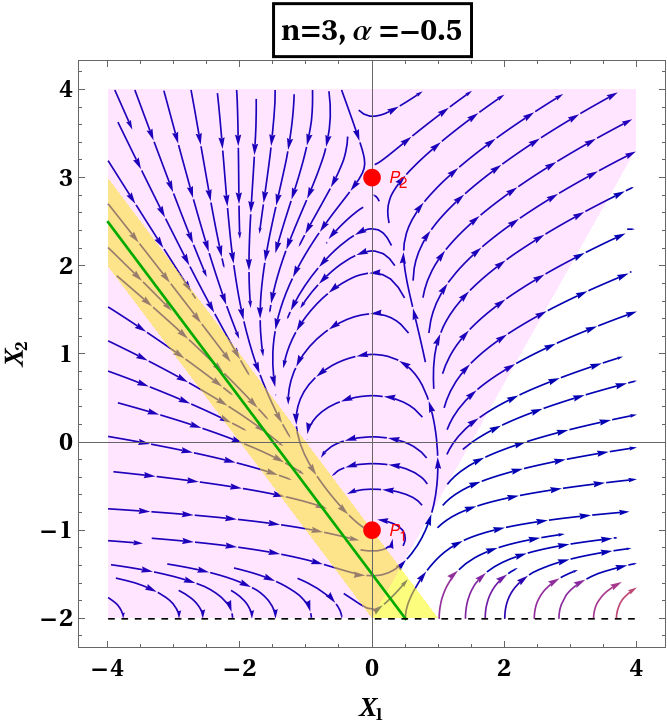}\hfill%
\includegraphics[width=.45\textwidth,height=8cm]{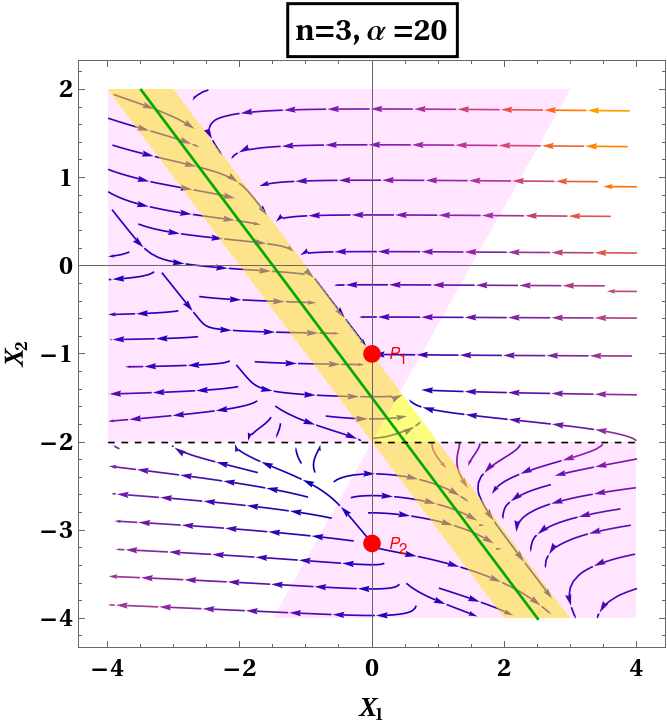}\\
\includegraphics[width=.45\textwidth,height=8cm]{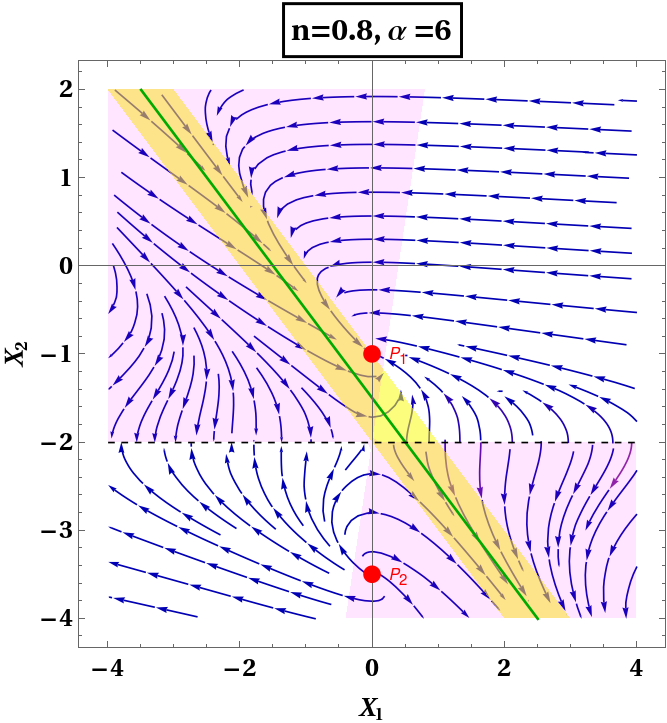}\hfill
\includegraphics[width=.45\textwidth,height=8cm]{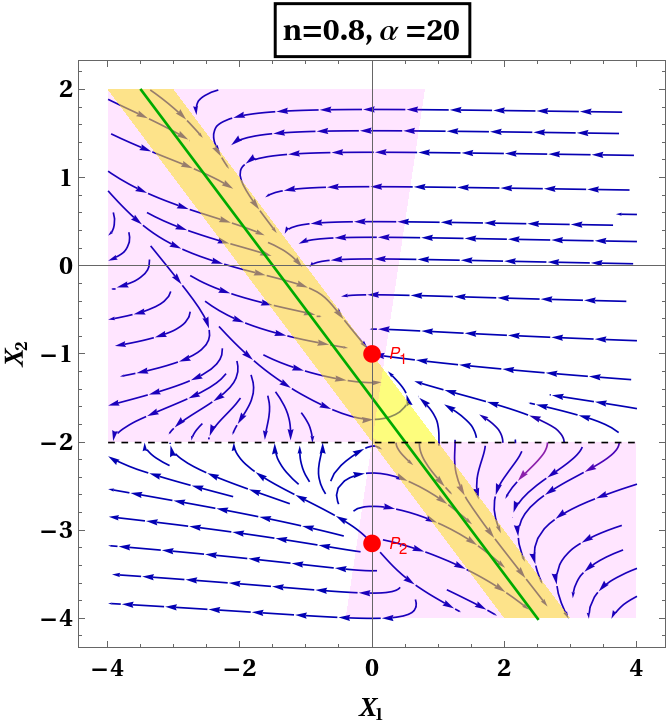}
\caption{Phase space trajectories of power-law model at different benchmark points. Critical points $P_1$ and $P_2$ are designated as a repeller and an attractor, respectively. The magenta region signifies positive jerk, and the yellow band represents $0 \leq \Omega_m \leq 1$ with $r_{\rm mc} \geq 0$. The central green line within the yellow band indicates the point where $r_{\rm mc}=1$.} 
\label{fig:B2}
\end{figure}   
 
   
\item  $P_{3\mp}$ and $P_{4\mp}$ Points: While both of these critical points can be defined in terms of model and coupling parameters, they both fall short of offering stable solutions during the de Sitter  phase. Additionally, they exhibit non-physical behavior in terms of critical matter density at the specific benchmark point chosen to analyze the phase-space trajectories of the 2-D system. Due to this non-physical characteristic observed in the density parameters at points $P_{3\mp}$ and $P_{4\mp}$ for the benchmark values of $n$ and $\alpha$, which have been chosen from the accessible range of $P_1$. We omitted these points from the phase space plot in fig.[\ref{fig:B2}].

\end{itemize}

\begin{figure}[H]
\centering
\includegraphics[width=.495\textwidth,height=5.5cm]{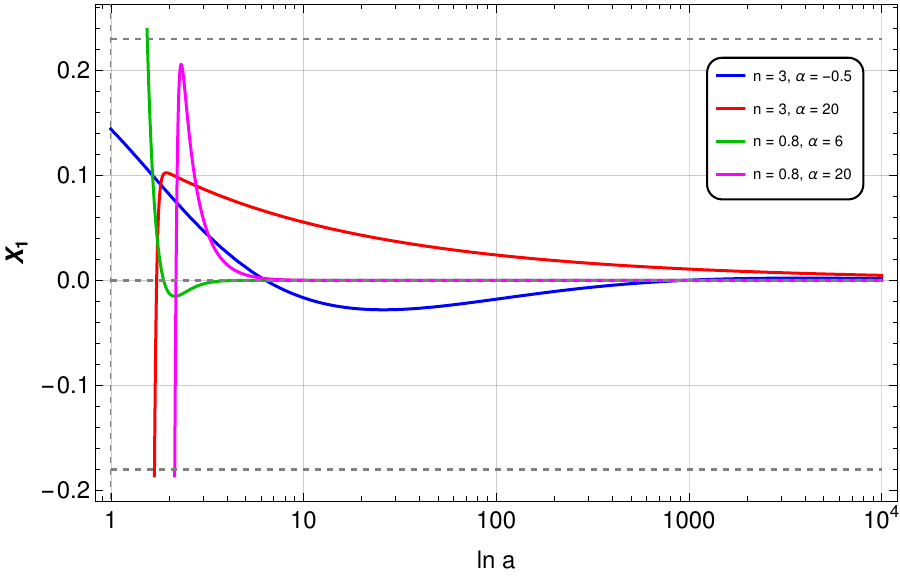}\hfill%
\includegraphics[width=.495\textwidth,height=5.5cm]{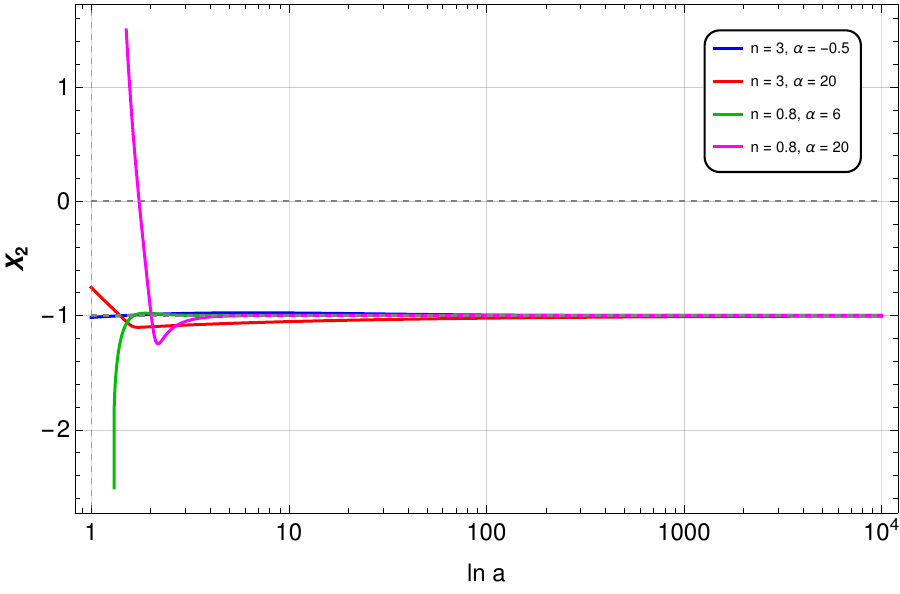}\hfill
\caption{ Evolution of $X_1$ and $X_2$ in power-law model at four benchmark values during the de Sitter phase.}
\label{fig:B3}
\end{figure}

 In fig.[\ref{fig:B3}], we have depicted the evolutionary dynamics of the dynamical variables $X_1$ and $X_2$ with respect to the logarithmic function of the FLRW scale factor ($a$), while maintaining the value of $X_3$ at 2, capturing their behavior during the de Sitter phase. The chosen model and coupling parameters fall within the acceptable range defined by $n$ and $\alpha$ for the power-law model, serving as benchmark points shown in fig.[\ref{fig:B1}]. Our primary focus is on examining the dynamic characteristics during the de Sitter phase, emphasizing the evolution from late-time to the distant future. Hence, our concern is not to showcase the highly unstable behavior of the dynamical variables during the early epochs. In the left panel, it is evident that the $X_1$ variable converges toward a value near zero, while in the right panel, the $X_2$ variable saturates near a value of negative unity. Since both of these dynamic variables rely entirely on the selected $f(R)$ models, the evolutionary plots provide insights into the models evolution near a critical point at specific benchmark values. Notably, both variables exhibit asymptotically stable behavior, indicating a stable acceleration in the late-time evolution with a specific choice of coupling and model parameters, all of which adhere to physical viability conditions.

\section{Conclusion}
\label{Sec:5}

In this article, we have investigated the potential incorporation of interactions between curvature and matter within the context of viable $f(R)$ dark-energy models. Our approach involves introducing a minimally coupled interacting term at the action level, allowing us to analyze the overall dynamics using a flat FLRW metric as the background. The selected form of interaction is designed to leave the total conserving continuity equation for the curvature and matter sectors unaffected. However, at the individual sector level, the presence of a source term ($\mathcal{Q}$) becomes crucial in influencing the dynamics. The chosen interaction term ($\mathcal{Q}$) effectively combines the influences of both matter and curvature sectors, determining the rate of energy transfer between them. This transfer rate reaches either a maximum or minimum based on the nature of the coupling strength ($\alpha$).\\

This study's crucial aspect involves simplifying the 3-D dynamical framework to a 2-D framework by keeping one of the $f(R)$ model-independent dynamical variables constant. Subsequently, it examines the dynamics of the curvature-matter interaction during the de Sitter  phase using the linear stability approach. The inclusion of the interaction term $\mathcal{Q}$ results in modifying the autonomous equations within the dynamical system compared to non-interacting scenarios. We identified critical points in the dynamical system and assessed their stability during the de Sitter  phase, separately considering two viable $f(R)$ models, namely, the generalized $\Lambda$-CDM model and the power-law model. The study thoroughly investigates how interaction parameters impact the stability characteristics of various fixed points within the dynamical system.\\

The introduction of a novel curvature-matter coupling in the context of $f(R)$ gravity results in a modification of critical points during the de Sitter  phase. Common acceptable regions, satisfying stability, critical matter density, and the ratio of matter to curvature density conditions, have been identified for both models based on coupling strength and model parameters. Different benchmark points within these regions have been selected to illustrate the behavior of critical points in 2-D phase-space plots in the presence of interaction. In each model, an attractor-type critical point has been identified that met all specified criteria. The nature of the attractor point has been categorized based on the $f(R)$ model. In the generalized $\Lambda$-CDM model, the attractor point depends on both model and coupling parameters, while in the power-law model, it is entirely independent of these parameters. For the stable attractor point in the generalized $\Lambda$-CDM model, the allowed coupling range is $\alpha<-2.15$, with the model parameter $bc$ ranging between 1.85 and 2.25, as seen in fig.[\ref{fig:A1}]. In contrast, for the power-law model, the acceptable coupling range is $\alpha>-1$ and the model parameter $n$ falls between $0.35-4$, as depicted in fig.[\ref{fig:B1}]. All these parameter ranges are accompanied by the de Sitter  phase.\\

Since our study primarily revolves around investigating matter-curvature interaction in viable $f(R)$ dark energy models during the de Sitter  phase. The constraint imposed by the grand equation of state (EoS) parameter during this phase leads to the value $\omega_{\rm tot.} = -1$. Eq.(\ref{eq:d8}) demonstrates that this specific condition results in a constant value for the dynamical variable $X_3 = 2$. Additionally, at this particular value of $X_3$, the deceleration parameter ($q$) is also determined to be $-1$. To understand the 2-D phase space behavior of $X_1$ and $X_2$ in this coupled system, we set the variable $X_3$ based on the behavior of the de Sitter phase in the presence of interacting curvature-matter interaction. This analysis is conducted for both the generalized $\Lambda$-CDM model and the power-law model. In the phase portrait of each model, we have included all critical points that exhibit physical behavior concerning critical matter density at selected benchmark values of model parameters ($bc$ or $n$) and coupling strength ($\alpha$). In the generalized $\Lambda$-CDM scenario, we have identified a model parameter-dependent critical point $P_2$ as an attractor, while in the power-law scenario, a model parameter-independent critical point $P_1$ has been identified as an attractor. Both attractors satisfy all physicality criteria and are situated within the shared region of magenta and yellow in the phase plot. The cosmological coincidence between the curvature-driven dark energy and dark matter is highlighted by the green line, where the energy densities of both sectors are equal. The attractor points are also near this line, indicating that the cosmological coincidence between dark energy and dark matter is achieved by introducing this specific form of interaction during the de Sitter  phase.  Evolution of dynamical variables $X_1$ and $X_2$ offers valuable insights into the behavior of the selected $f(R)$ gravity model at permissible benchmark points (three benchmark values for generalized $\Lambda$-CDM model and four for power-law model). Given our emphasis on exploring this study within the de Sitter phase of the universe, our focus lies specifically on investigating the dynamical variables ($X_1$ and $X_2$) behavior from late-time to the distant future, rather than dwelling on the highly unstable early epochs. Both variables for both models have demonstrated asymptotic stability near the critical points, utilizing the chosen values of the benchmark points.\\

  In conclusion, this study advances our comprehension of the intricate dynamics of interacting curvature-matter scenarios within the framework of modified $f(R)$ gravity theories. The identification of stable attractors during the de Sitter  phase, nuanced stability features, and phase trajectory patterns contribute to a comprehensive understanding of the complex behavior exhibited by the system. The insights presented not only deepen our knowledge of the dynamics in these systems but also provide new opportunities for exploration in cosmology, offering valuable perspectives into the fundamental dynamics of the universe.

\paragraph{Acknowledgement}\
Author would like to thank the referess for their valuable comments. Special thanks are extended to Abhijit Bandyopadhyay and Rahul Roy for their enriching and insightful discussions.

\end{document}